\documentstyle[prbbib,aps,floats,epsf]{revtex} 
 
\begin{document}
 
\draft                    
 
\wideabs{
 
 
\title{
         \vskip -0.5cm
         \hfill\hfil{\rm\normalsize Printed on \today}\\
         Electronic and Structural Properties of Carbon Nano-Horns}
 
\author{ Savas Berber,
         Young-Kyun Kwon
         and David Tom\'anek }
 
\address{ Department of Physics and Astronomy, and
         Center for Fundamental Materials Research, \\
         Michigan State University,
         East Lansing, Michigan 48824-1116}
 
\date{Received \hspace*{3.0cm}}
 
\maketitle
 
 
\begin{abstract}
We use parametrized linear combination of atomic orbitals
calculations to determine the stability, optimum geometry and
electronic properties of nanometer-sized capped graphitic cones,
called ``nano-horns''. Different nano-horn morphologies are
considered, which differ in the relative location of the five
terminating pentagons. Simulated scanning tunneling microscopy images
of the various structures at different bias voltages reflect a net
electron transfer towards the pentagon vertex sites. We find that the
local density of states at the tip, observable by scanning tunneling
spectroscopy, can be used to discriminate between different tip
structures. Our molecular dynamics simulations indicate that
disintegration of nano-horns at high temperatures starts in the
highest-strain region near the tip.
\end{abstract}
 
 
}
 
 
\narrowtext
 
Since their first discovery \cite{Iijima91}, carbon nanotubes have
drawn the attention of both scientists and engineers due to the large
number of interesting new phenomena they exhibit
\cite{{Dresselhaus96},{Ebbesen},{Yakobson},{RSaito98}}, and due to
their potential use in nanoscale devices: quantum wires
\cite{Dekker97}, nonlinear electronic elements \cite{Zettl97},
transistors \cite{Dekker-Nature98}, molecular memory devices
\cite{PRLxNTM}, and electron field emitters
\cite{{emitter1},{emitter2},{emitter3},{emitter4}}. Even though
nanotubes have not yet found commercially viable applications,
projections indicate that this should occur in the very near future,
with the advent of molecular electronics and further miniaturization
of micro-electromechanical devices (MEMS). Among the most unique
features of nanotubes are their electronic properties. It has been
predicted that single-wall carbon nanotubes
\cite{{Iijima93},{Bethune93}} can be either metallic or
semiconducting, depending on their diameter and chirality
\cite{{Mintmire92},{RSaito92},{Oshiyama92}}. Recently, the
correlation between the chirality and conducting behavior of nanotubes
has been confirmed by high resolution scanning tunneling microscopy
(STM) studies \cite{{Dekker98},{Lieber98}}.
 
Even though these studies have demonstrated that atomic resolution
can be achieved \cite{{Dekker98},{Lieber98},{Johnson98}}, the precise
determination of the {\em atomic configuration}, characterized by the
chiral vector, diameter, distortion, and position of atomic defects,
is still a very difficult task to achieve in nanotubes. Much of the
difficulty arises from the fact that the electronic states at the
Fermi level are only indirectly related to the atomic positions.
Theoretical modeling of STM images has been found crucial to correctly
interpret experimental data for graphite
\cite{{Tomanek-STM1},{Tomanek-STM2}}, and has been recently applied to
carbon nanotubes \cite{{Meunier99},{Mele99}}. As an alternative
technique, scanning tunneling spectroscopy combined with modeling has
been used to investigate the effect of the terminating cap on the
electronic structure of nanotubes \cite{Kim99}.
 
Among the more unusual systems that have been synthesized in the past
few years are cone-shaped graphitic carbon structures
\cite{{cone1},{cone}}. Whereas similar structures have been observed
previously near the end of multi-wall nanotubes \cite{horn92}, it is
only recently that an unusually high production rate of up to
$10$~g/h has been achieved for single-walled cone-shaped structures,
called ``nano-horns'', using the CO$_2$ laser ablation technique at
room temperature in absence of a metal catalyst \cite{horn}. These
conical nano-horns have the unique opening angle of
${\approx}20^{\circ}$.
 
We consider a microscopic understanding of the electronic and
structural properties of nano-horns a crucial prerequisite for
understanding the role of terminating caps in the physical behavior
of contacts between nanotube-based nano-devices. So far, neither
nano-horns nor other cone-shaped structures have been investigated
theoretically. In the following, we study the structural stability of
the various tip morphologies, and the inter-relationship between the
atomic arrangement and the electronic structure at the terminating
cap, as well as the disintegration behavior of nano-horns at high
temperatures.
 
Cones can be formed by cutting a wedge from planar graphite and
connecting the exposed edges in a seamless manner. The opening angle
of the wedge, called the disclination angle, is $n(\pi/3)$, with
$0{\le}n{\le}6$. This disclination angle is related to the opening
angle of the cone by $\theta=2\sin^{-1}(1-n/6)$. Two-dimensional
planar structures (e.g. a graphene sheet) are associated with $n=0$,
and one-dimensional cylindrical structures, such as the nanotubes, are
described by $n=6$. All other possible graphitic cone structures with
$0<n<6$ have been observed in a sample generated by pyrolysis of
hydrocarbons \cite{cone}. According to Euler's rule, the terminating
cap of a cone with the disclination angle $n(\pi/3)$ contains $n$
pentagon(s) that substitute for the hexagonal rings of planar
graphite.
 
The observed cone opening angle of ${\approx}20^{\circ}$,
corresponding to a $5\pi/3$ disclination, implies that all nano-horns
contain exactly five pentagons near the tip. We classify the
structure of nano-horns by distinguishing the relative positions of
the carbon pentagons at the apex which determine the morphology of the
terminating cap. Our study will focus on the influence of the relative
position of these five pentagons on the properties of nano-horns.
 
\begin{figure}
    \centerline{
        \epsfxsize=0.3\columnwidth
        \epsffile{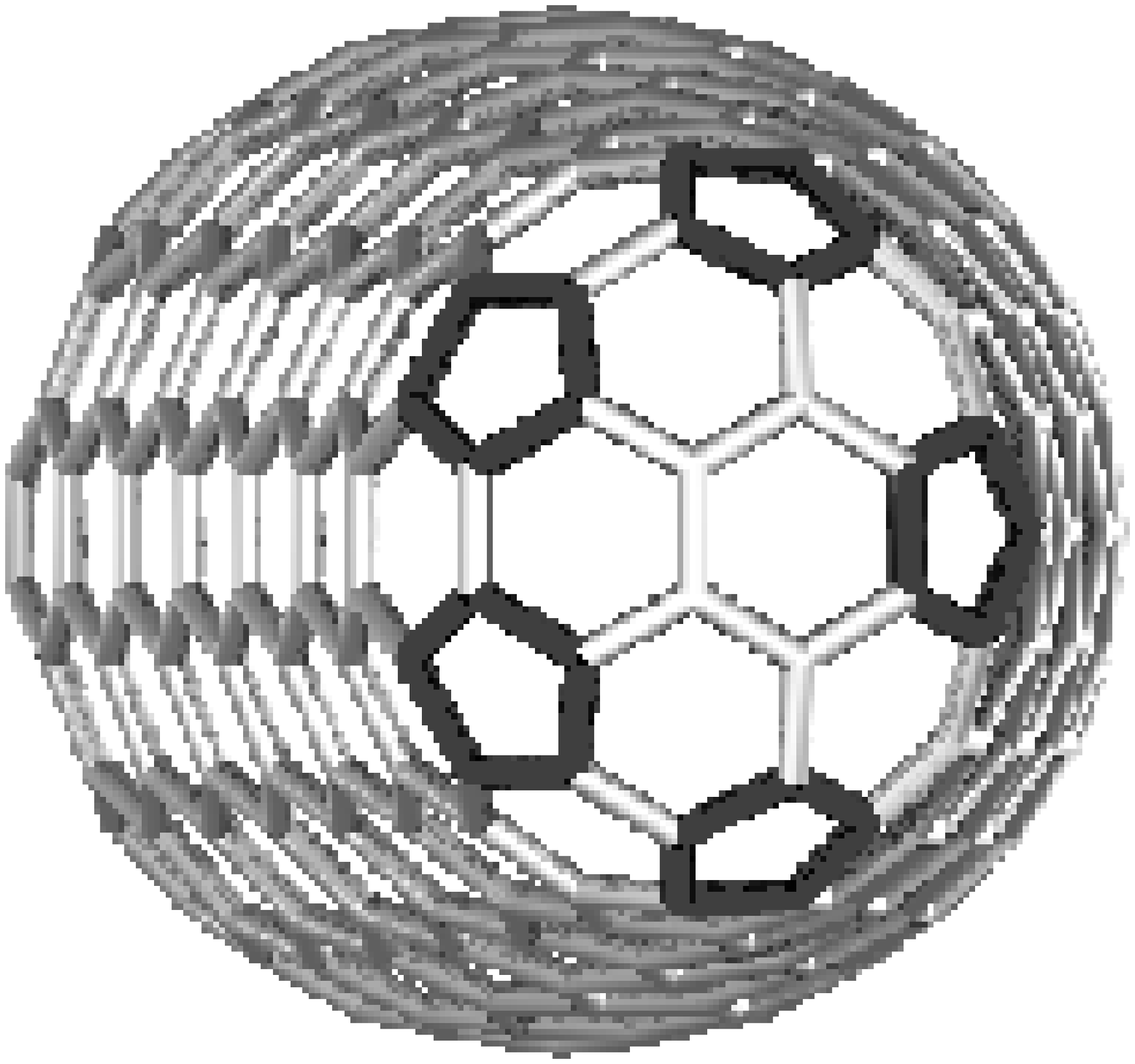}
        \hspace*{0.005\columnwidth}
        \epsfxsize=0.3\columnwidth
        \epsffile{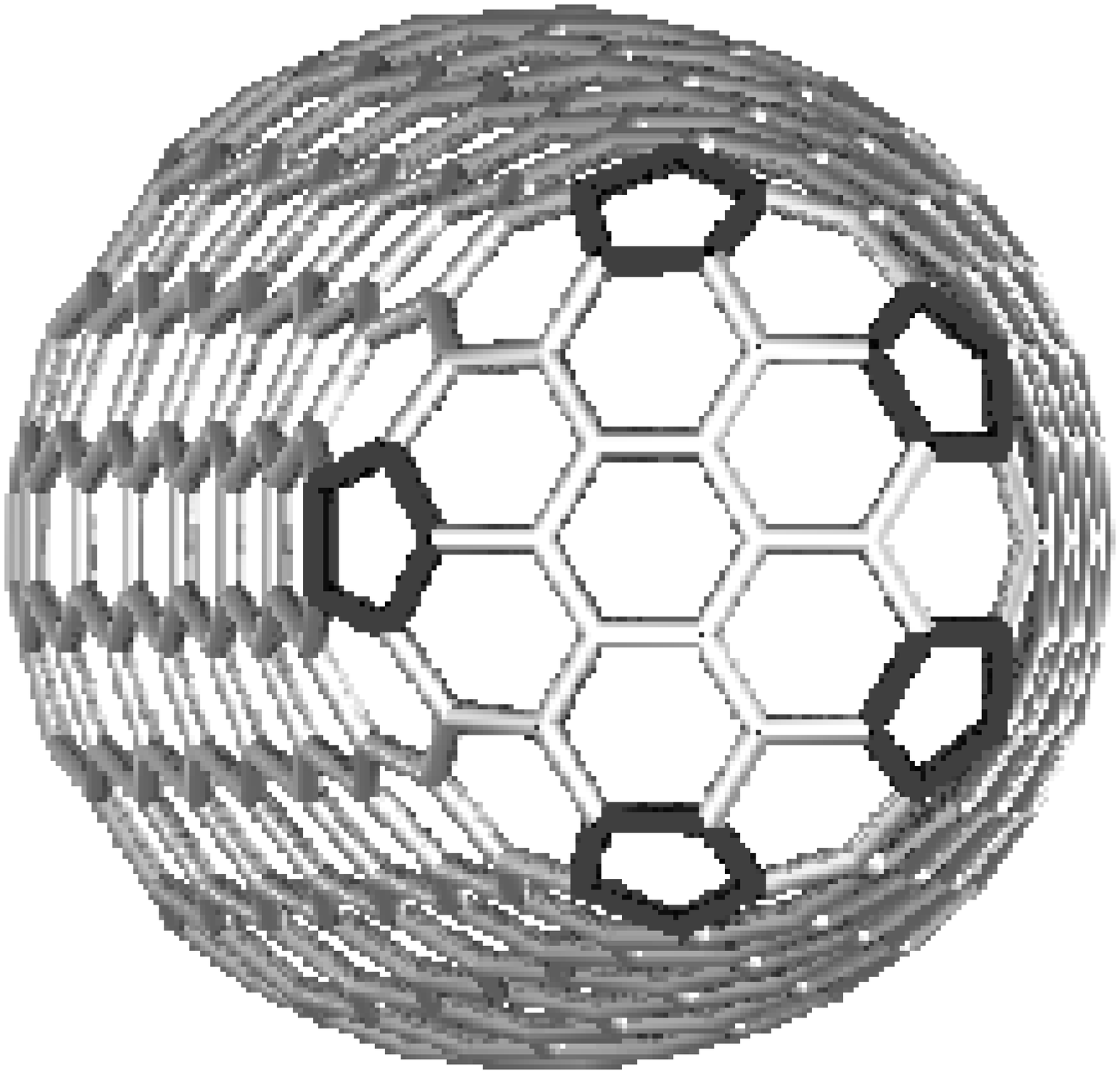}
        \hspace*{0.005\columnwidth}
        \epsfxsize=0.3\columnwidth
        \epsffile{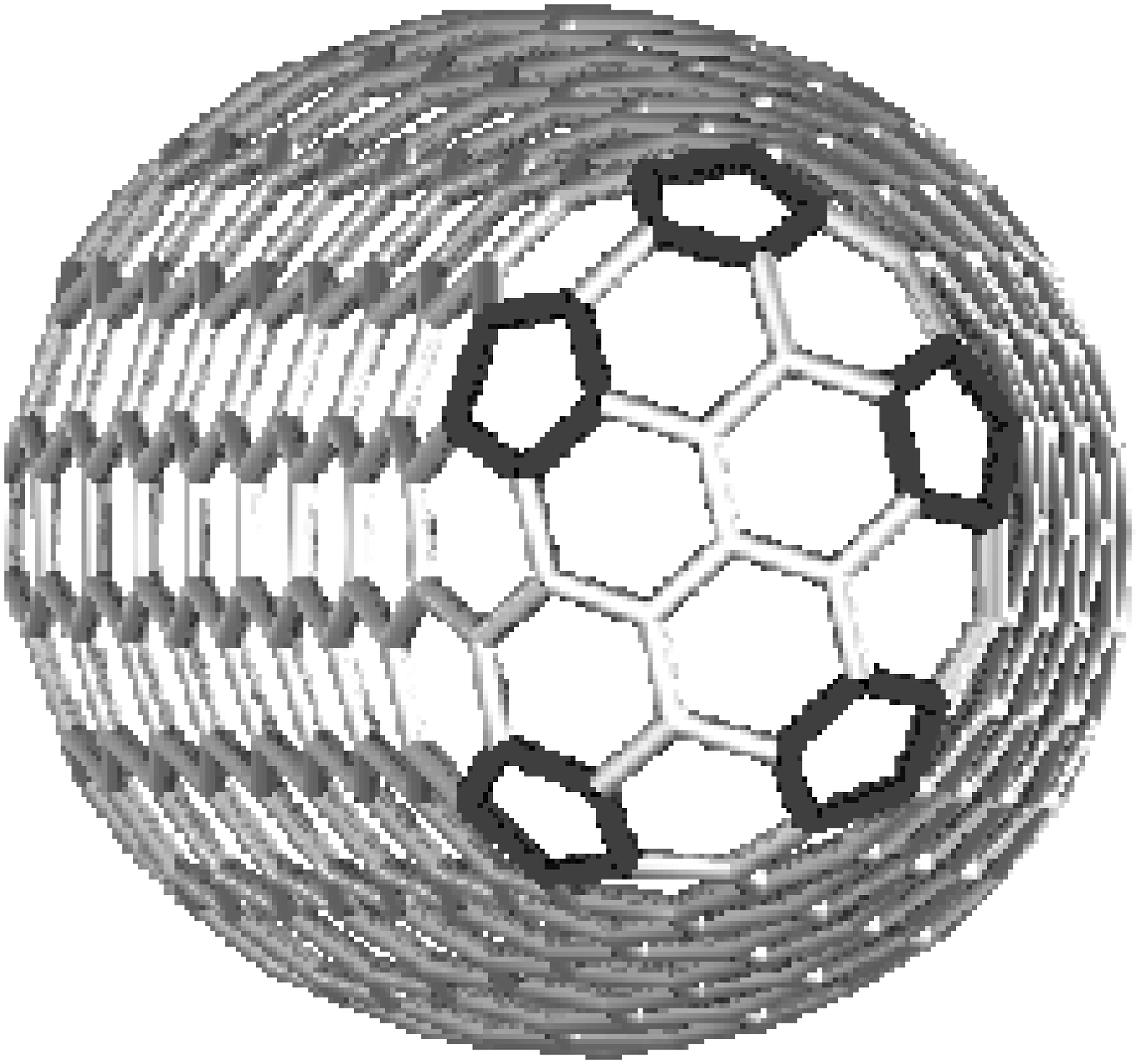}
    }
    \vspace*{-0.3\columnwidth}
    \centerline{
        \hspace*{-0.30\columnwidth}
        {\large\bf (a)}
        \hspace*{0.22\columnwidth}
        {\large\bf (b)}
        \hspace*{0.24\columnwidth}
        {\large\bf (c)}
    }
    \vspace*{0.25\columnwidth}
    \centerline{
        \epsfxsize=0.3\columnwidth
        \epsffile{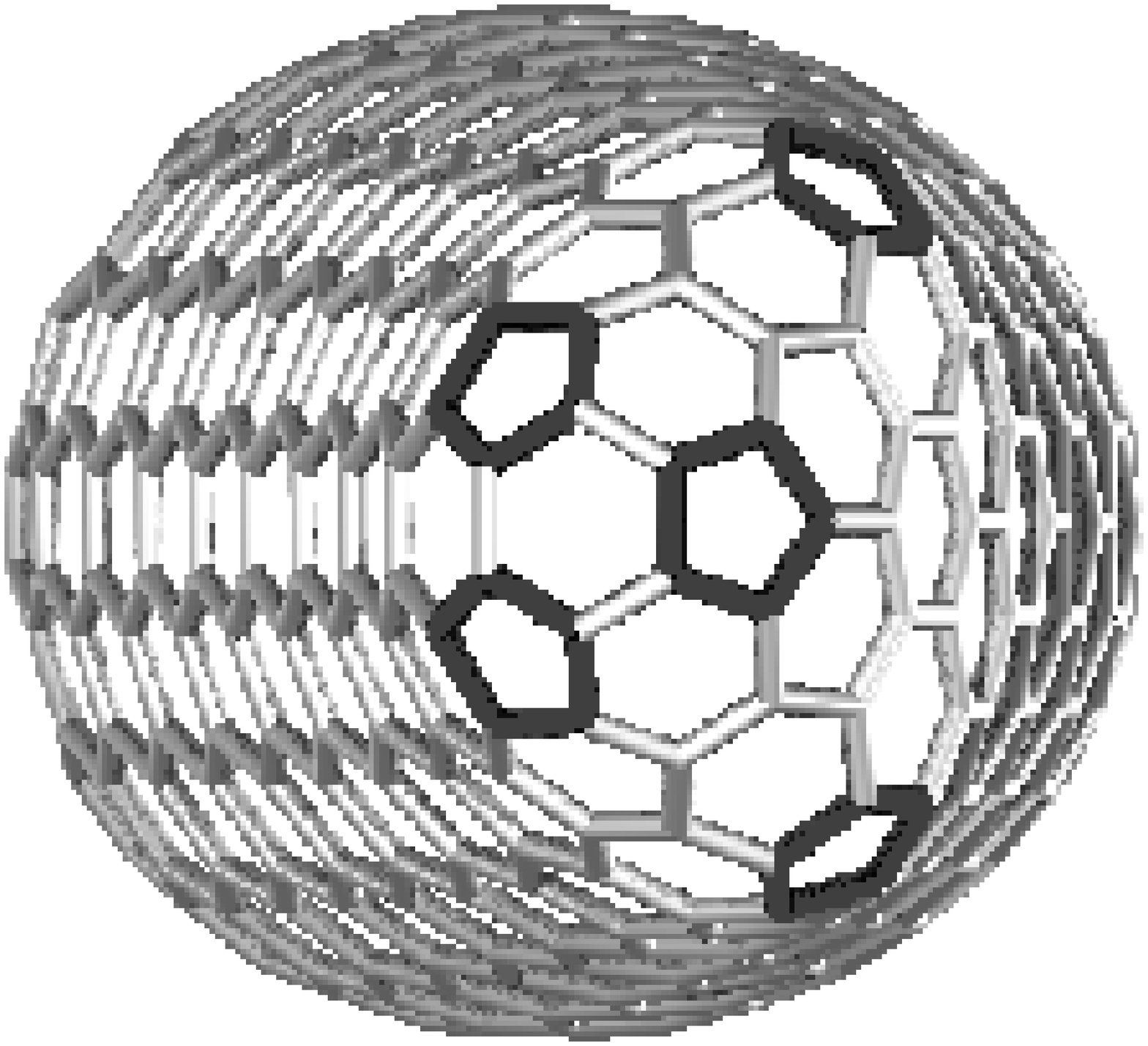}
        \hspace*{0.005\columnwidth}
        \epsfxsize=0.3\columnwidth
        \epsffile{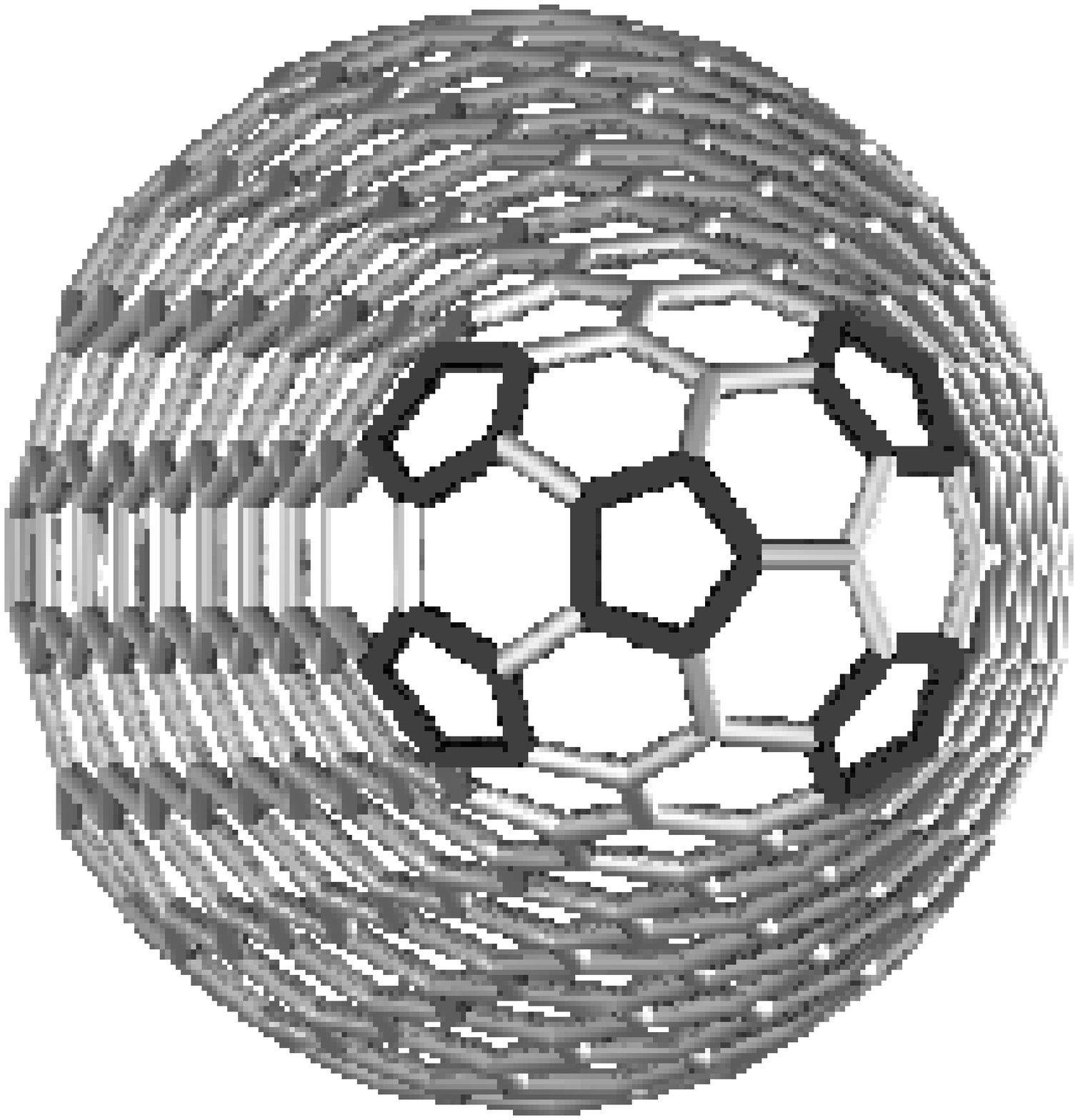}
        \hspace*{0.005\columnwidth}
        \epsfxsize=0.3\columnwidth
        \epsffile{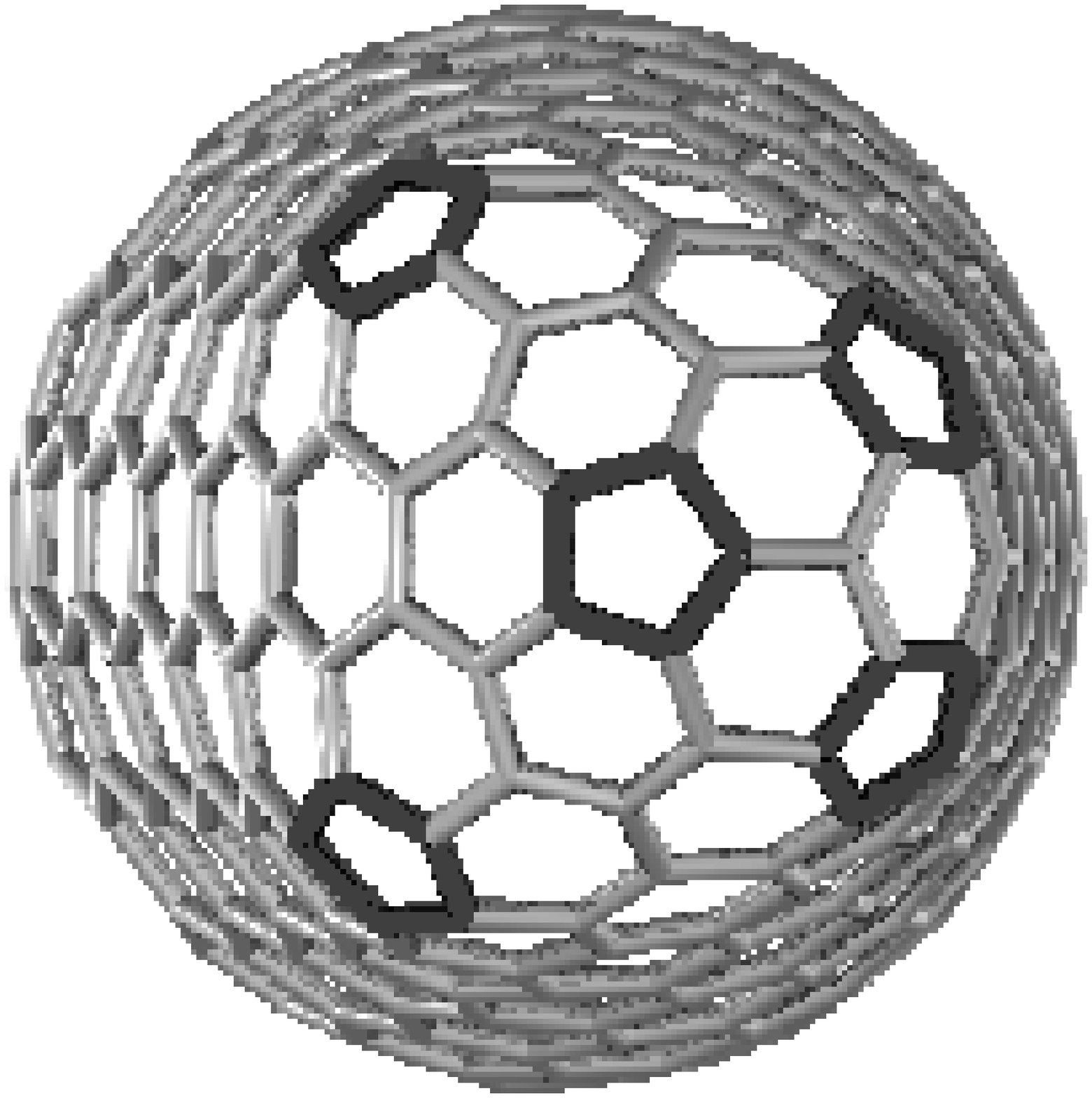}
    }
    \vspace*{-0.3\columnwidth}
    \centerline{
        \hspace*{-0.30\columnwidth}
        {\large\bf (d)}
        \hspace*{0.22\columnwidth}
        {\large\bf (e)}
        \hspace*{0.24\columnwidth}
        {\large\bf (f)}
    }
    \vspace*{0.25\columnwidth}
\caption{
Optimized carbon nano-horn structures with a total disclination
angle of $5(\pi/3)$, containing five isolated pentagons at the
terminating cap. Structures (a) -- (c) contain all pentagons at
the conical ``shoulder'', whereas structures (d) -- (f) contain
a pentagon at the apex. The pentagons are highlighted by a
darker color.
}
\label{Fig1}
\end{figure}
 
The cap morphologies investigated in this study are presented in
Fig.~\ref{Fig1}. Nano-horns with all five pentagons at the
``shoulder'' of the cone, yielding a blunt tip, are shown in
Figs.~\ref{Fig1}(a)--(c). Nano-horns with a pentagon at the
apex of the tip, surrounded by the other four pentagons at the
shoulder, are shown in Figs.~\ref{Fig1}(d)--(f). Note that
the cone angle of each nano-horn is ${\approx}20^{\circ}$, even
though the size of the terminating cap varies with the relative
position of the pentagons.
 
To determine the structural and electronic properties of carbon
nano-horns, we used the parametrized linear combination of atomic
orbitals (LCAO) technique with parameters determined by {\em ab
initio} calculations for simpler structures \cite{PRLxCAR}. This
approach has been found useful to describe minute electronic structure
and total energy differences for systems with too large unit cells to
handle accurately by {\em ab initio} techniques. Some of the problems
tackled successfully by this technique are the electronic structure
and superconducting properties of the doped C$_{60}$ solid
\cite{C60-supercond}, the opening of pseudo-gaps near the Fermi level
in a (10,10) nanotubes rope \cite{{PRLxNTR},{JMRxNTR}} and a
(5,5)@(10,10) double-wall nanotube \cite{PRLxDWT}, as well as
fractional quantum conductance in nanotubes \cite{PRLxTSP}. This
technique, combined with the recursion technique to achieve an $O(N)$
scaling, can determine very efficiently the forces on individual
atoms \cite{SSCxRT}, and had previously been used with success to
describe the disintegration dynamics of fullerenes \cite{PRLxMLT},
the growth of multi-wall nanotubes \cite{PRLxLIP} and the dynamics of
a ``bucky-shuttle'' \cite{PRLxNTM}.
 
To investigate the structural stability and electronic properties of
carbon nano-horns, we first optimized the structures with various cap
morphologies, shown in Fig.~\ref{Fig1}. For the sake of an easier
interpretation of our results, we distinguish the $N_{\rm
cap}{\approx}40-50$ atoms at the terminating cap from those within the
cone-shaped mantle, that is terminated by $N_{\rm edge}$ atoms at the
other end. We associate the tip region of a hypothetically infinite
nano-horn with all the sites excluding the edge. Structural
details and the results of our stability calculations are presented
in Table~\ref{Table1}. These results indicate that atoms in
nano-horns are only ${\approx}0.1$~eV less stable than in graphite.
The relative differences in $<E_{\rm coh,tot}>$ reflect the strain
energy changes induced by the different pentagon arrangements. To
minimize the effect of under-coordinated atoms at the edge on the
relative stabilities, we excluded the edge atoms from the average when
calculating $<E_{\rm coh,tip}>$. Since our results for $<E_{\rm
coh,tip}>$ and $<E_{\rm coh,tot}>$ follow the same trends, we believe
that the effect of edge atoms on the physical properties can be
neglected for structures containing hundreds of atoms. Even though
the energy differences may appear minute on a per-atom basis, they
translate into few electron-volts when related to the entire
structure. Our results suggest that the under-coordinated edge atoms
are all less stable than the cone mantle atoms by ${\approx}0.5$~eV.
Also atoms in pentagons are less stable than those in hexagons by
${\approx}0.1$~eV, resulting in an energy penalty of
${\approx}0.5$~eV to create a pentagon if the strain energy induced by
bending the lattice could be ignored.
 
\begin{table}
\setdec 0.000
\caption{
Structural data and stability results for carbon nano-horn
structures (a)--(f), presented in Fig.~\protect\ref{Fig1}.
$N_{\rm tot}=N_{\rm tip}+N_{\rm edge}$ is the total number of
atoms, which are subdivided into tip and edge atoms. $<E_{\rm
coh,tot}>$ is the average binding energy, taken over the entire
structure, and $<E_{\rm coh,tip}>$ the corresponding value
excluding the edge region. $<E_{\rm coh,edge}>$ is the binding
energy of the edge atoms, and $<E_{\rm coh,pent}>$ is the
average over the pentagon sites in each system.
}
\begin{tabular}{lcccccc}
Quantity      & (a)   & (b)   &  (c)   &  (d)  &  (e)  &  (f) \\
\tableline
$N_{\rm tot}$ & $205$ & $272$ & $ 296$ & $290$ & $308$ & $217$ \\
$N_{\rm tip}$ & $172$ & $233$ & $ 257$ & $251$ & $270$ & $180$ \\
$N_{\rm edge}$ & $33$ & $39$ & $39$ & $39$ & $38$ & $37$ \\
$<E_{\rm coh,tot}>$~(eV) & $-7.28$ & $-7.29$ & $-7.30$ & $-7.30$ &
                                               $-7.31$ & $-7.28$ \\
$<E_{\rm coh,tip}>$~(eV) & $-7.36$ & $-7.36$ & $-7.37$ & $-7.36$ &
                                               $-7.37$ & $-7.36$ \\
$<E_{\rm coh,edge}>$~(eV) & $-6.88$ & $-6.88$ & $-6.88$ &
                            $-6.88$ & $-6.87$ & $-6.89$ \\
$<E_{\rm coh,pent}>$~(eV) & $-7.28$ & $-7.28$ & $-7.28$ &
                            $-7.28$ & $-7.28$ & $-7.28$ \\
\end{tabular}
\label{Table1}
\end{table}

When comparing the stabilities of the tip regions, described by
$<E_{\rm coh,tip}>$, we found no large difference between blunt tips
that have all the pentagons distributed along the cylinder mantle and
pointed tips containing a pentagon at the apex. We found the
structure shown in Fig.~\ref{Fig1}(c) to be more stable than the
other blunt structures with no pentagon at the apex. Similarly, the
structure shown in Fig.~\ref{Fig1}(e) is most stable among the
pointed tips containing a pentagon at the apex. Equilibrium
carbon-carbon bond lengths in the cap region are
$d_{CC}=1.43-1.44$~{\AA} at the pentagonal sites and
$d_{CC}=1.39$~{\AA} at the hexagonal sites, as compared to
$d_{CC}=1.41-1.42$~{\AA} in the mantle. This implies that the
``single bonds'' found in pentagons should be weaker than the
``double bonds'' connecting hexagonal sites, thus confirming our
results in Table~\ref{Table1} and the analogous behavior in the
C$_{60}$ molecule.
 
\begin{figure}
    \centerline{
        \epsfxsize=0.45\columnwidth
        \epsffile{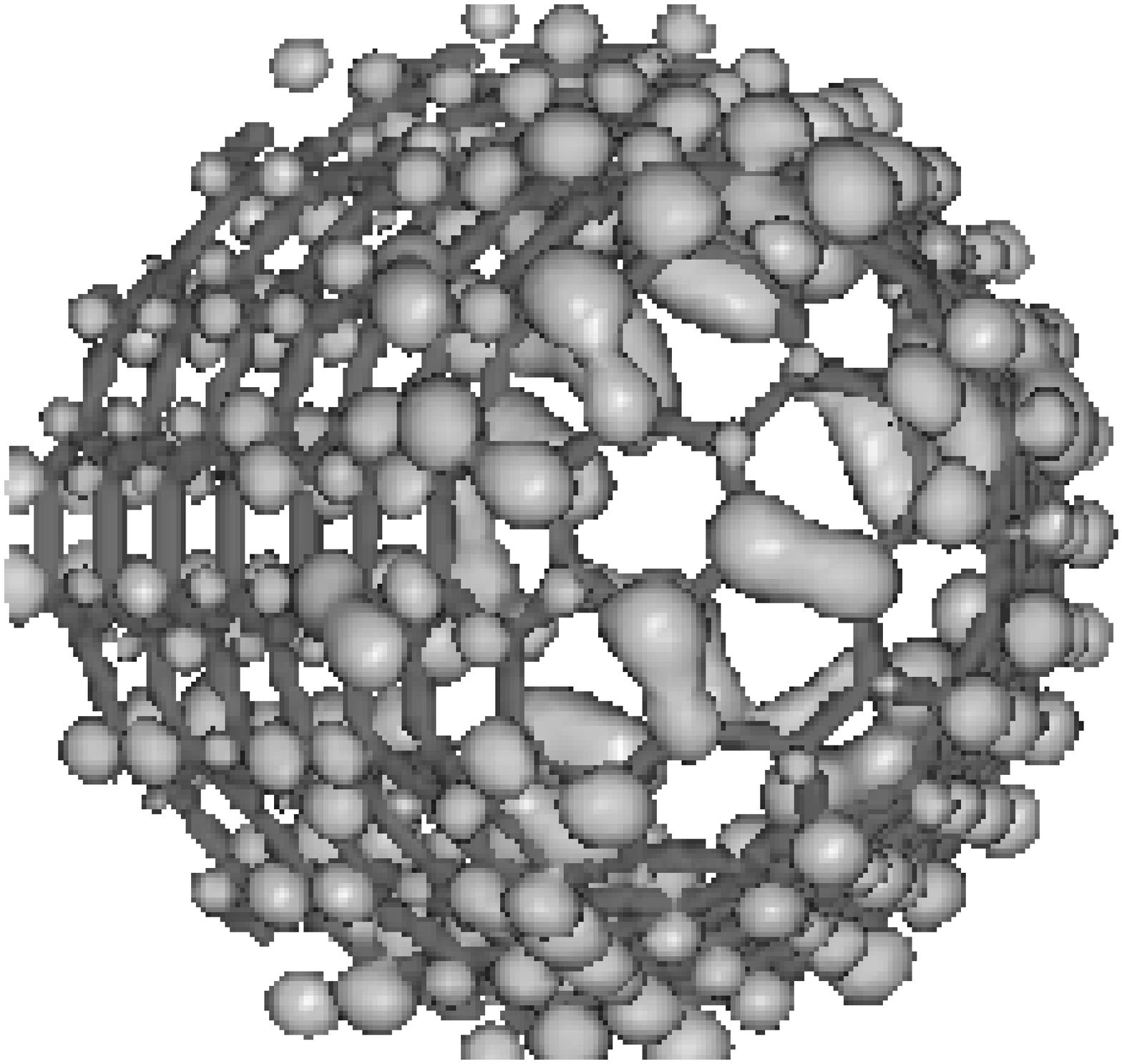}
        \epsfxsize=0.45\columnwidth
        \epsffile{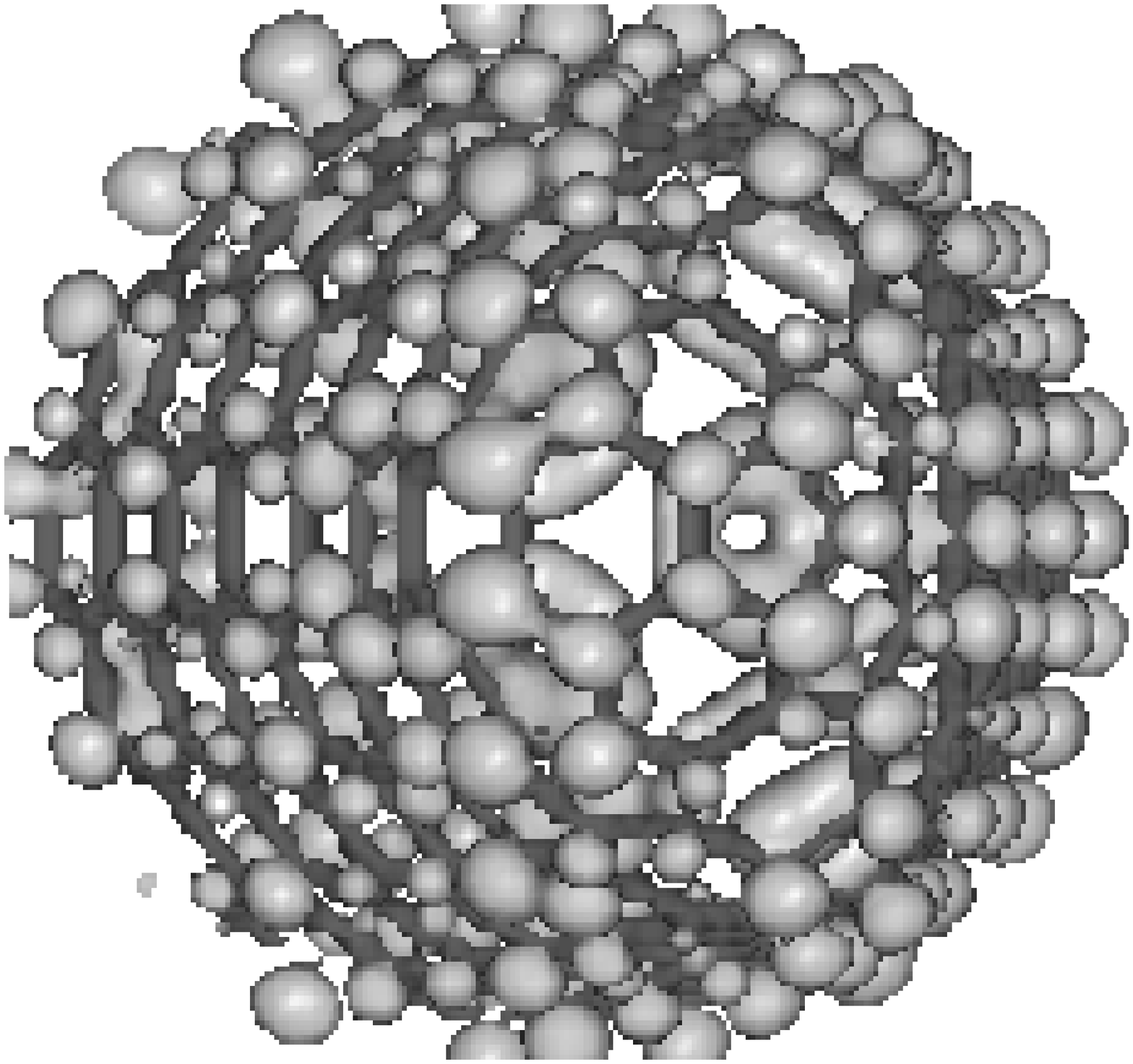}
    }
    \vspace*{-0.45\columnwidth}
    \centerline{
        \hspace*{-0.3\columnwidth}
        {\large\bf (a)}
        \hspace*{0.35\columnwidth}
        {\large\bf (b)}
    }
    \vspace*{0.40\columnwidth}
\caption{
Simulated STM images in the tip region for
(a) the nano-horn shown in Fig.~\protect\ref{Fig1}(c) and
(b) the nano-horn shown in Fig.~\protect\ref{Fig1}(d).
These results for the occupied electronic states near the Fermi
level, corresponding to the bias voltage of $V_b=0.2$~V, are
suggestive of a net electron transfer from the hexagonal to the
pentagonal sites. The charge density contour displayed
corresponds to the value of
$\rho=1.35{\times}10^{-3}$~electrons/{\AA}$^3$. Dark lines
depict the atomic bonds to guide the eye.
}
\label{Fig2}
\end{figure}
 
Since pentagon sites are defects in an all-hexagon structure, they
may carry a net charge. To characterize the nature of the defect
states associated with these sites, we calculated the electronic
structure at the tip of the nano-horns. The charge density associated
with states near $E_F$, corresponding to the local density of states
at that particular position and energy, is proportional to the
current observed in STM experiments. To compute the local charge
density associated with a given eigenstate, we projected this state
onto a local atomic basis. The projection coefficients were used in
conjunction with real-space atomic wave functions from density
functional calculations \cite{Tomanek-STM2} to determine the charge
density corresponding to a particular level or the total charge
density. To mimic a large structure, we convoluted the discrete level
spectrum by a Gaussian with a full-width at half-maximum of $0.3$~eV.
Using this convoluted spectrum, we also determined the charge density
associated with particular energy intervals corresponding to STM data
for a given bias voltage.
 
In Fig.~\ref{Fig2}, we present such simulated STM images for the
nano-horns represented in Figs.~\ref{Fig1}(c) and \ref{Fig1}(d). We
show the charge density associated with occupied states within a
narrow energy interval of $0.2$~eV below the Fermi level \cite{Fermi}
as three-dimensional charge density contours, for the density value of
$\rho=1.35\times10^{-3}$~electrons/{\AA}$^3$. Very similar results to
those presented in Fig.~\ref{Fig2} were obtained at a higher bias
voltage of $0.4$~eV. As seen in Fig.~\ref{Fig2}, $pp\pi$ interactions
dominate the spectrum near $E_F$. These images also show a net excess
charge on the pentagonal sites as compared to the hexagonal sites.
This extra negative charge at the apex should make pointed nano-horn
structures with a pentagon at the apex better candidates for field
emitters \cite{{emitter1},{emitter2},{emitter3},{emitter4}} than
structures with no pentagon at the apex and a relatively blunt tip.
 
\begin{figure}
    \centerline{
        \epsfxsize=0.6\columnwidth
        \epsffile{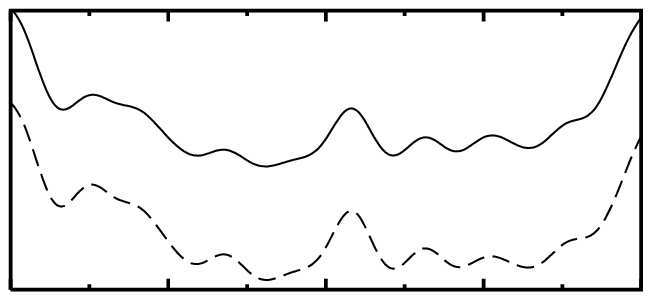}
        \hspace*{-0.12\columnwidth}
        \epsfxsize=0.6\columnwidth
        \epsffile{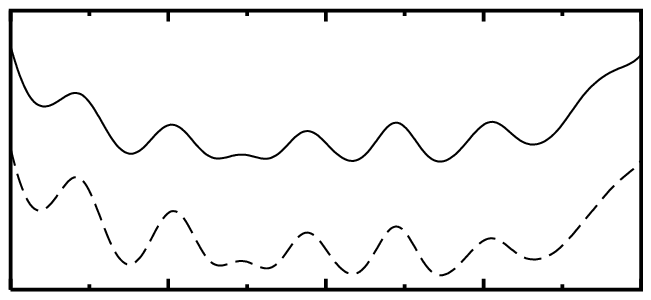}
    }
    \vspace*{-0.19\columnwidth}
    \centerline{
        \hspace*{-0.3\columnwidth}
        {\large\bf (a)}
        \hspace*{0.395\columnwidth}
        {\large\bf (d)}
    }
    \vspace*{-0.1\columnwidth}
    \centerline{
        \epsfxsize=0.6\columnwidth
        \epsffile{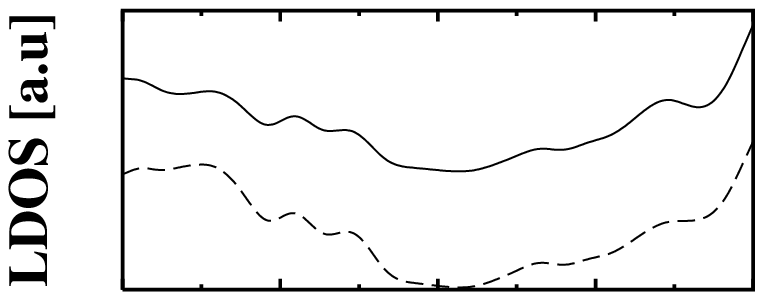}
        \hspace*{-0.12\columnwidth}
        \epsfxsize=0.6\columnwidth
        \epsffile{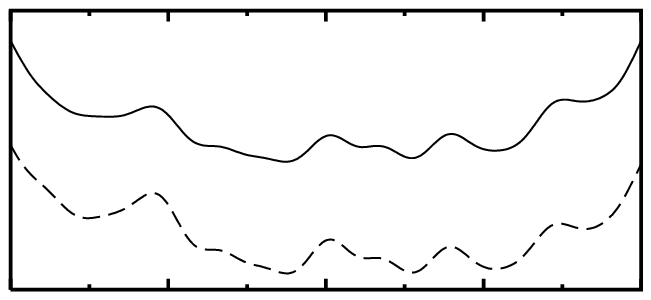}
    }
    \vspace*{-0.19\columnwidth}
    \centerline{
        \hspace*{-0.3\columnwidth}
        {\large\bf (b)}
        \hspace*{0.395\columnwidth}
        {\large\bf (e)}
    }
    \vspace*{-0.1\columnwidth}
    \centerline{
        \epsfxsize=0.6\columnwidth
        \epsffile{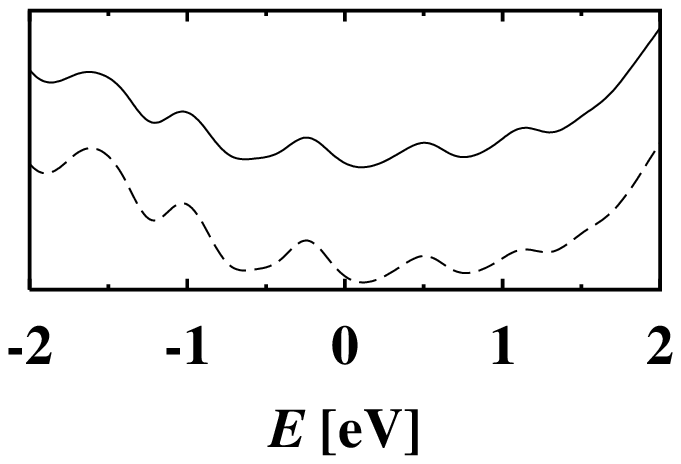}
        \hspace*{-0.12\columnwidth}
        \epsfxsize=0.6\columnwidth
        \epsffile{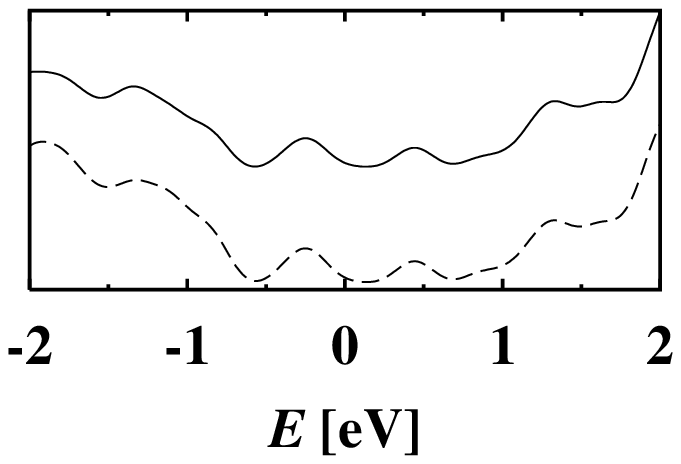}
    }
    \vspace*{-0.19\columnwidth}
    \centerline{
        \hspace*{-0.3\columnwidth}
        {\large\bf (c)}
        \hspace*{0.395\columnwidth}
        {\large\bf (f)}
    }
    \vspace*{0.15\columnwidth}
\caption{
Local electronic densities of states, normalized per atom and
shown in consistent arbitrary units, for the terminating cap of
the tip (solid line) and for the pentagon sites only (dashed
line) of nano-horn structures shown in Fig.~\protect\ref{Fig1}.
The similarity between the results for the entire cap and the
pentagon sites necessitated a vertical offset of one set of data
for an easy distinction, thus indicating that variations in the
arrangement of pentagons affect the densities of states not only
at the pentagonal sites, but within the entire cap region. The
most marked differences are noted near the Fermi level, at
$E{\approx}0$~eV.
}
\label{Fig3}
\end{figure}
 
It has been shown previously that theoretical modeling of STM images
is essential for the correct interpretation of experimental data.
Atomically resolved STM images, however, are very hard to obtain
especially near the terminating caps of tubes \cite{Kim99} and cones,
due to the large surface curvature that can not be probed efficiently
using current cone-shaped STM tips. A better way to identify the tip
structure may consist of scanning tunneling spectroscopy (STS)
measurements in the vicinity of the tip. This approach is based on
the fact that in STS experiments, the normalized conductance
$(V/I)(dI/dV)$ is proportional to the local density of states which,
in turn, is structure sensitive. We have calculated the local density
of states at the terminating cap of the tips for the different
nano-horn structures shown in Fig.~\ref{Fig1}. Our results are shown
in Fig.~\ref{Fig3}, convoluted using a Gaussian with a full-width at
half-maximum of $0.3$~eV.
 
To investigate the effect of pentagonal sites on the electronic
structure at the tip, we first calculated the local density of states
only at the $25$ atoms contained in the five terminating pentagons.
The corresponding densities of states, shown by the dashed curve in
Fig.~\ref{Fig3}, are found to vary significantly from structure to
structure near the Fermi level \cite{Fermi}. Thus, a comparison
between the densities of states at $E{\approx}E_F$ should offer a new
way to discriminate between the various tip morphologies. For an easy
comparison with experiments, we also calculated the local density of
states in the entire terminating cap, including all five pentagons
and consisting of $N{\approx}40-50$ atoms, depending on the
structure. The corresponding density of states, given by the solid
line in Fig.~\ref{Fig3}, is vertically displaced for easier
comparison. Our results show that the densities of states, both
normalized per atom, are very similar. Thus, we conclude that the
pentagonal sites determine all essential features of the electronic
structure near the Fermi level at the tip.
 
Next, we have studied the heat resilience \cite{horn} as well as the
decay mechanism of nano-horns at extremely high temperatures using
molecular dynamics simulations \cite{Tildesley}. In our canonical
molecular dynamics simulations, we keep the structure at a constant
temperature using a Nos\'e-Hoover thermostat \cite{Nose-Hoover}, and
use a fifth--order Runge--Kutta interpolation scheme to integrate the
equations of motion, with a time step of $\Delta t = 5\times
10^{-16}$~s. We found the system to remain structurally intact within
the temperature range from $T=2,000-4,000$~K. Then, we heated up the
system gradually from $T=4,000$~K to $5,000$~K within 4,000 time
steps, corresponding to a time interval of $2$~ps. Our molecular
simulations show that nano-horn structures are extremely heat
resilient up to $T{\alt}4,500$~K. At higher temperatures, we find
these structures to disintegrate preferentially in the vicinity of
the pentagon sites. A simultaneous disintegration of the nano-horn
structures at the exposed edge, which also occurs in our simulations,
is ignored as an artifact of finite-size systems. The preferential
disintegration in the higher strain region near the pentagon sites,
associated with a large local curvature, is one reason for the
observation that nano-horn tips are opened easily at high
temperatures, in presence of oxygen \cite{horn}.
 
In summary, we used parametrized linear combination of atomic orbitals
calculations to determine the stability, optimum geometry and
electronic properties of nanometer-sized capped graphitic cones,
called nano-horns. We considered different nano-horn morphologies
that differ in the relative location of the five terminating
pentagons. We found a net electron transfer to the pentagonal sites
of the cap. This extra charge is seen in simulated scanning tunneling
microscopy images of the various structures at different bias
voltages. We found that the local density of states at the tip,
observable by scanning tunneling spectroscopy, can be used to
discriminate between different tip structures. Our molecular dynamics
simulations indicate that disintegration of nano-horns at high
temperatures starts in the highest-strain region near the tip.
 
We thank S. Iijima for fruitful discussions on carbon nano-horns. We
acknowledge financial support by the Office of Naval Research and
DARPA under Grant Number N00014-99-1-0252.
 


\begin{references}
\bibitem{Fermi} The nano-horn structures considered in our calculation
	 are clusters containing edge atoms. We found that due to
         the charge transfer towards the under-coordinated atoms at the
         edge, the ``Fermi level'' of these clusters may drop by
         as much as ${\approx}1.0-1.2$~eV with respect to that of a
         graphene monolayer. To compensate for this artifact, we
         excluded the edge region from the calculated density of states
         and found the $E_F$ value to lie very close to that of a
         graphene monolayer.
\end{references}
\end{document}